\documentclass[aps,prc,twocolumn,superscriptaddress,showpacs,floatfix,nofootinbib]{revtex4}
\usepackage{amsmath,graphicx,ulem}

\begin{document}

\title{Thermal noise in a boost-invariant matter expansion in relativistic heavy-ion collisions}

\author{Chandrodoy Chattopadhyay}
\affiliation{Department of Nuclear and Atomic Physics, Tata Institute of
Fundamental Research, Homi Bhabha Road, Mumbai 400005, India}
\author{Rajeev S. Bhalerao}
\affiliation{Department of Physics, Indian Institute of Science Education and Research (IISER), 
Homi Bhabha Road, Pune 411008, India}
\author{Subrata Pal}
\affiliation{Department of Nuclear and Atomic Physics, Tata Institute of
Fundamental Research, Homi Bhabha Road, Mumbai 400005, India}

\begin{abstract}
We formulate a general theory of thermal fluctuations within causal second-order viscous hydrodynamic 
evolution of matter formed in relativistic heavy-ion collisions. The fluctuation is treated perturbatively 
on top of a boost-invariant longitudinal expansion. Numerical simulation of thermal noise is performed 
for a lattice QCD equation of state and for various second-order dissipative evolution equations.
Phenomenological effects of thermal fluctuations on the two-particle rapidity correlations  
are studied.

\end{abstract}

\pacs{25.75.Ld, 24.10.Nz, 47.75+f}


\maketitle

\section{Introduction}

Relativistic dissipative hydrodynamics has become an important tool to
study bulk properties of the near-equilibrium system formed in
relativistic heavy-ion collisions
\cite{Adams:2005dq,Adcox:2004mh,ALICE:2011ab,ATLAS:2012at,Chatrchyan:2013kba}.
The large densities and short mean-free times in the system allow for
a coarse graining in hydrodynamics which integrates out the microscopic
length and time scales. The effective degrees of freedom are then the
average conserved quantities, namely the energy, momentum, electric
charge and baryon number, which are dynamically evolved according to
the hydrodynamic equations
\cite{Muller:1967zza,Israel:1979wp,Muronga:2003ta,Romatschke:2007mq,Song:2007ux}.
When low enough densities are reached and the interaction times become
longer, the system falls out of equilibrium, which ultimately leads to
a kinetic freeze-out.

In spite of this inherent coarse graining \cite{Romatschke:2009im},
hydrodynamics has been remarkably successful in explaining several
experimental observables pertaining to relativistic heavy-ion
collisions, for example, the anisotropic flow $v_n$
\cite{Ollitrault:1992bk}, that characterizes the final-state momentum
anisotropy in the plane transverse to the beam direction. The flow has
been well understood as the collective hydrodynamic response to the
initial collision geometry fluctuating event by event
\cite{Alver:2010gr,Schenke:2011bn,Qiu:2011hf,Gale:2012rq,Bhalerao:2015iya,Chattopadhyay:2017bjs}.
The long-range rapidity structures observed in multiparticle correlation
measurements in heavy-ion collisions \cite{Chatrchyan:2013nka}
as well as in high-multiplicity
collision events involving small projectiles ($p/d/^3$He) 
 \cite{Khachatryan:2015lva,Khachatryan:2016txc}
can also be
related to the hydrodynamic behavior.

It is then instructive to investigate whether the thermal noise or the
fast microscopic degrees of freedom that survive coarse graining, have
any measurable effect on the experimental observables. The
fluctuation-dissipation theorem already forces one to consider
fluctuations in systems that are in thermal equilibrium. Further, as
the size of the fireball formed is just about 10 fm, and there are
only a finite number of particles in each coarse-grained fluid cell,
fluctuations may play a crucial role. Thermal noise could be even more
important for proper interpretation of observables near the critical
point for confinement-deconfinement transition where all fluctuations
are large in general. In contrast to the perturbation created in the
medium due to energy deposition by a propagating jet
\cite{Staig:2010pn}, thermal fluctuations are produced at all
space-time points in the fluid cells. These local fluctuations are
propagated/diffused via the fluid dynamic evolution
equations. Nevertheless, the thermal fluctuations in heavy-ion
collisions may not be quite large, other than near the critical point,
as the strongly coupled quark-gluon plasma (QGP) is formed with a
small shear viscosity to entropy density of $\eta_v/s \simeq 0.08-0.20$
\cite{Schenke:2011bn,Qiu:2011hf,Gale:2012rq,Bhalerao:2015iya,Chattopadhyay:2017bjs}.

The theory of hydrodynamic fluctuations or noise in a nonrelativistic
fluid \cite{Landau} was extended into the relativistic regime for
Navier-Stokes (first-order) viscous fluid by Kapusta {\it et al.}
\cite{Kapusta:2011gt}. The thermal fluctuation, $\Xi^{\mu\nu}$, of the
energy-momentum tensor was shown to have a nontrivial autocorrelation
$\langle \Xi^{\mu\nu}(x) \Xi^{\alpha\beta}(x') \rangle \sim T \eta_v
\: \delta^4(x-x')$, where $T$ is the temperature.
\cite{Kapusta:2011gt,Young:2013fka,Nagai:2016wyx}. Due to the
occurrence of the Dirac delta function, the energy and momentum
density averaged value of this white noise becomes $\sim
1/\sqrt{\Delta V \Delta t}$. Thus, even for small shear viscosity
$\eta_v$, the white noise sets a lower limit on the system cell size
$\Delta V$ that is essentially comparable to the correlation
length. Consequently, white noise could lead to large gradients which
makes the basic hydrodynamic formulation (based on gradient expansion)
questionable. The divergence problem can be circumvented, by treating
the white noise as a perturbation (in a linearized hydrodynamic
framework) on top of a baseline nonfluctuating hydrodynamic evolution
\cite{Kapusta:2011gt,Young:2013fka,Young:2014pka}. Analytic solutions
for hydrodynamic fluctuations were obtained in the case of
boost-invariant longitudinal expansion without transverse dynamics
(Bjorken flow) \cite{Kapusta:2011gt} and with transverse dynamics
(Gubser flow) \cite{Yan:2015lfa}. However, both these calculations
were performed in the relativistic Navier-Stokes theory for a
conformal fluid.

It is important to recall that the first-order dissipative fluid
dynamics or the Navier-Stokes theory, displays acausal behavior that
may lead to unphysical effects.  On the other hand, the second-order
dissipative fluid dynamics, based on the M\"uller-Israel-Stewart (MIS)
framework
\cite{Muller:1967zza,Israel:1979wp,Muronga:2003ta,Romatschke:2009im},
gives hyperbolic equations and restores causality. The commonly used
MIS formulation has been quite successful in explaining the spectra
and azimuthal anisotropy of particles produced in heavy-ion
collisions. Recently, formally new dissipative equations have been
derived from Chapman-Enskog-like iterative expansion of the Boltzmann
equation in the relaxation-time approximation
\cite{Jaiswal:2013npa,Bhalerao:2013pza} and from the modified
14-moment method which was developed by Denicol {\it et al.}
\cite{Denicol:2012cn}.

In this work, we shall present the formulation of thermal fluctuations
for these forms of hydrodynamic dissipative evolution equations in the
case of a boost-invariant longitudinal expansion. The fluctuation
equations so obtained are rather general and will be used along with
an equation of state (EOS) corresponding to a conformal fluid and then
with the lattice QCD EOS. Since analytical solutions for hydrodynamic
fluctuations cannot be obtained for the baseline second-order
hydrodynamic approaches, we shall perform numerical simulations of
thermal noise and its evolution as a perturbation on top of
boost-invariant longitudinal expansion of the viscous medium.

The paper is organized as follows. In Sec. II, we formulate
hydrodynamic fluctuations in the linearized limit as a perturbation on
top of second-order dissipative equations for boost-invariant
expansion. In Sec. III, we calculate the phenomenology of particle
freeze-out and the effect of fluctuation on two-particle rapidity
correlations. In Sec. IV, we present results from numerical
simulations for the rapidity correlations with ideal gas and lattice
QCD equations of state. Finally in Sec. V, we summarize our results
and conclude.

\section{Thermal fluctuations in boost invariant hydrodynamics}

In this section we formulate thermal
fluctuations in the boost-invariant longitudinal expansion of matter
within second-order viscous hydrodynamics. In the presence
of a thermal noise tensor $\Xi^{\mu\nu}$, the total energy-momentum
tensor becomes
\begin{align}\label{FTt:eq}
T^{\mu\nu} = \epsilon u^\mu u^\nu - p\Delta^{\mu\nu} + \pi^{\mu\nu} + \Xi^{\mu\nu}.
\end{align}
We shall work in the Landau-Lifshitz frame and disregard particle
diffusion current, which is a reasonable
approximation due to very small values of the net baryon number formed
at RHIC and LHC; we further ignore bulk viscosity in our calculation.
In the above equation, $\epsilon$ and $p$ are respectively the energy
density and pressure in the fluid's local rest frame (LRF), and
$\pi^{\mu\nu}$ is the shear pressure
tensor. $\Delta^{\mu\nu}=g^{\mu\nu}-u^\mu u^\nu$ is the projection
operator on the three-space orthogonal to the hydrodynamic
four-velocity $u^\mu$ which is defined
by the Landau matching condition $T^{\mu\nu}u_{\nu} = \epsilon u^\mu$.

The total energy-momentum tensor $T^{\mu\nu}$ consists of an average
part $T_0^{\mu\nu}$ (represented by subscript ``0") and a thermally
fluctuating part $\delta T^{\mu\nu}$ (represented by
$\delta$). In the presence of fluctuations,
the energy density (or temperature), flow velocity, and shear pressure
tensor can be written as \cite{Kapusta:2011gt}
\begin{align}\label{nonlin:eq}
\epsilon &=  \epsilon_0 + \delta\epsilon, \nonumber\\
u^\mu &=  u_0^\mu + \delta u^\mu,  \nonumber\\
\pi^{\mu\nu} &=  \pi_0^{\mu\nu} + \delta\pi^{\mu\nu}.  
\end{align}
In the linearized limit (keeping terms up to first order in fluctuations), the total energy-momentum tensor becomes: 
\begin{align}\label{FT:eq}
T^{\mu\nu} &= \epsilon u^\mu u^\nu - p\Delta^{\mu\nu} + \pi^{\mu\nu} + \Xi^{\mu\nu}, \nonumber\\  
&= T_0^{\mu\nu} + \delta T_{\rm id}^{\mu\nu} + \delta\pi^{\mu\nu} + \Xi^{\mu\nu}
\equiv T_0^{\mu\nu} + \delta T^{\mu\nu}. 
\end{align}
The total fluctuating part $\delta T^{\mu\nu}$ has contributions from 
the viscous term $\delta\pi^{\mu\nu}$, the noise term $\Xi^{\mu\nu}$, and the
ideal energy-momentum tensor term 
\begin{align}\label{DTid:eq}
\delta T_{\rm id}^{\mu\nu} =& \delta\epsilon \: u_0^\mu u_0^\nu - \delta p \: \Delta^{\mu\nu}_0 \nonumber\\ 
& + \left(\epsilon_0 + p_0 \right)\left( u_0^\mu \delta u^\nu + \delta u^\mu u_0^\nu \right)
+ {\cal O}(\delta^2). 
\end{align}
All of these can be determined by the fluctuating variables 
($\delta\epsilon, \delta u^\mu, \delta \pi^{\mu\nu}$). 
The conservation equations for the total energy-momentum tensor, $\partial_\mu T^{\mu\nu}=0$,
along with the usual conservation for the average part, $\partial_\mu T_0^{\mu\nu}=0$, lead to
\begin{align}\label{FCT:eq}
\partial_\mu (\delta T_{\rm id}^{\mu\nu} + \delta\pi^{\mu\nu} + \Xi^{\mu\nu})
\equiv \partial_\mu (\delta T^{\mu\nu}) = 0.
\end{align}
Though in a single event thermal noise causes $\delta T^{\mu\nu} \neq 0$, the average over many events 
results in $\langle \delta T^{\mu\nu} \rangle = 0$. However, noise induces a 
nonvanishing correlator $\langle \delta T^{\mu\nu} \delta T^{\alpha\beta} \rangle$, 
that contributes to event-by-event distribution of an observable, e.g., 
two-particle rapidity correlations \cite{Kapusta:2011gt,Young:2014pka}.
In order to solve (numerically) Eq. (\ref{FCT:eq}), 
which involves the evolution 
of viscous fluctuation and noise, one requires the averaged 
quantities ($\epsilon_0, u_0^\mu, \pi_0^{\mu\nu}$).  We first deal with 
the background viscous evolution equations and then formulate the 
evolution of fluctuation on top this background.

For Bjorken longitudinal expansion, we work in the Milne coordinates ($\tau,x,y,\eta$) 
where proper time $\tau = \sqrt{t^2-z^2}$, space-time rapidity $\eta = \ln[(t+z)/(t-z)]/2$,
and four-velocity $u_0^\mu = (1,0,0,0)$.
The conservation equation for the average part of the energy-momentum tensor, 
$\partial_\mu T_{0\: id}^{\mu\nu} = -\partial_\mu \pi_0^{\mu\nu}$,
gives the evolution equation of noiseless $\epsilon_0$ as 
\begin{align}\label{energy:eq}
\frac{d\epsilon_0}{d\tau} = -\frac{1}{\tau} \left(\epsilon_0 + p_0 - \pi_0 \right),
\end{align}
where $\tau^2\pi_0^{\eta\eta} \equiv -\pi_0$ is taken as the independent component 
of the shear pressure tensor. 
For the three independent variables, we need two more equations, namely, the viscous
evolution equation and the equation of state. 
The simplest choice for the dissipative equation would be the relativistic 
Navier-Stokes theory, where the instantaneous constituent equation for the shear 
pressure is 
$\pi^{\mu\nu} = 2\eta_v \nabla^{\langle \mu}u^{\nu\rangle} \equiv 2\eta_v \sigma^{\mu\nu}$.
Using Eq. (\ref{nonlin:eq}), the average (noiseless) shear part in the Bjorken case becomes
\begin{align}\label{NS:eq}
\pi_0 = \frac{4\eta_v}{3}\theta_0 ,
\end{align}
where $\eta_v \geq 0$ is the shear viscosity coefficient, and 
$\nabla^{\langle \mu}u^{\nu\rangle} = (\nabla^\mu u^\nu + \nabla^\nu u^\mu)/2 - 
(\nabla \cdot u) \Delta^{\mu\nu}/3$ and $\nabla^\mu = \Delta^{\mu\nu}\partial_\nu$.
For boost-invariant case, the local expansion rate and the time derivative in the LRF are
$\theta_0 = 1/\tau$. In the following we shall use the standard notation
$A^{\langle\mu\nu\rangle}\equiv \Delta^{\mu\nu}_{\alpha\beta}A^{\alpha\beta}$ for traceless 
symmetric projection orthogonal to $u^{\mu}$, where 
$\Delta^{\mu\nu}_{\alpha\beta} \equiv (\Delta^\mu_\alpha \Delta^\nu_\beta +
\Delta^\mu_\beta \Delta^\nu_\alpha)/2 - (1/3)\Delta^{\mu\nu}\Delta_{\alpha\beta}$. 

The most commonly used second-order dissipative hydrodynamic equation derived from positivity 
of the divergence of entropy four-current is based on the works of M\"uller-Israel-Stewart (MIS) 
\cite{Muller:1967zza,Israel:1979wp,Muronga:2003ta,Romatschke:2007mq}:
\begin{align}\label{IS:eq}
\dot\pi^{\langle \mu\nu \rangle} =& - \frac{1}{\tau_\pi} \left( \pi^{\mu\nu} 
- 2\eta_v \nabla^{\langle \mu}u^{\nu\rangle} \right) \nonumber \\
& -\frac{1}{2} \pi^{\mu\nu} \frac{\eta_v T}{\tau_\pi} \partial_{\lambda} 
\left( \frac{\tau_\pi}{\eta_v T} u^\lambda \right) ,
\end{align}
where the above equation involves the full hydrodynamic variables, and 
$T = T_0 + \delta T$ is the total temperature corresponding to $\epsilon$.
In contrast to the first-order equation, the above equation restores causality 
by enforcing the shear pressure to relax to its first-order value via the 
relaxation time $\tau_\pi = 2\eta_v \beta_2$, where $\beta_2$ is the 
second-order transport coefficient. In the boost-invariant scaling expansion, 
the noiseless part of the dissipative Eq. (\ref{IS:eq}) reduces to 
\begin{align}\label{MIS:eq}
\frac{d\pi_0}{d\tau} + \frac{\pi_0}{\tau_\pi} 
= \frac{4\eta_v}{3\tau_\pi} \theta_0 - \lambda_\pi\theta_0\pi_0,
\end{align}
where terms up to second-order in the velocity gradients are kept in the expansion
of the last term in Eq. (\ref{IS:eq}). The resulting expansion coefficient is 
\begin{align}\label{lpi:eq}
\lambda_\pi = \frac{1}{2} \left[ 1 + \frac{\epsilon_0 + p_0}{T_0} \frac{dT_0}{d\epsilon_0} 
\left(1 - \frac{T_0}{\beta_2} \frac{d\beta_2}{dT_0} \right) \right],
\end{align}
which for an ultrarelativistic EOS reduces to $\lambda_\pi = 4/3$. 
The relaxation time $\tau_\pi$ depends on the underlying microscopic theory 
namely, weakly coupled QCD, lattice QCD, and ${\cal N}=4$ SYM \cite{Romatschke:2009im}.
For all these theories, one can express $\tau_\pi = \chi \eta_v/(sT_0)$, 
where the coefficient $2 \lesssim \chi \lesssim 6$.
In the present study we consider $\tau_\pi = 2\eta_v \beta_2 = 5\eta_v/(sT_0)$.
Hereafter $\eta_v/s$ is kept fixed, where $s = s_0 + \delta s$ is the total entropy density
in the linearized limit with $s_0$ being the average entropy density.

Alternatively, dissipative evolution equations can be derived from
Chapman-Enskog-like (CE) gradient expansion of the 
nonequilibrium distribution function about the local value, using Knudsen number 
as a small expansion parameter \cite{Jaiswal:2013npa,Bhalerao:2013pza,Chattopadhyay:2014lya}. 
The relativistic Boltzmann equation, in the relaxation-time approximation for the 
collision term, can be solved iteratively to yield 
\begin{equation}\label{SOSHEAR:eq}
\dot{\pi}^{\langle\mu\nu\rangle} \!+ \frac{\pi^{\mu\nu}}{\tau_\pi}\!= 
\frac{\sigma^{\mu\nu}}{\beta_2} \!+2\pi_\gamma^{\langle\mu}\omega^{\nu\rangle\gamma}
\!-\frac{10}{7}\pi_\gamma^{\langle\mu}\sigma^{\nu\rangle\gamma} 
\!-\frac{4}{3}\pi^{\mu\nu}\theta,
\end{equation}
where $\omega^{\mu\nu}\equiv(\nabla^\mu u^\nu-\nabla^\nu u^\mu)/2$ 
is the vorticity tensor.
In the boost-invariant case, the noiseless part of Eq. (\ref{SOSHEAR:eq}) gives
\begin{align}\label{CE:eq}
\frac{d\pi_0}{d\tau} + \frac{\pi_0}{\tau_\pi} 
= \frac{4\eta_v}{3\tau_\pi} \theta_0 - \lambda_\pi\theta_0\pi_0.
\end{align}
In the Chapman-Enskog-like approach, the relaxation time naturally comes out to be
$\tau_\pi = 2\eta_v\beta_2 = 5\eta_v/(s T_0)$ and $\lambda_\pi = 38/21$ \cite{Jaiswal:2013npa}.
In this limit the CE equation is equivalent to that obtained by Denicol et. al. 
\cite{Denicol:2012cn} where the expansion is controlled by the Knudsen number and the
inverse Reynolds number. We shall explore the effects of the above viscous equations on 
the thermal noise correlators and the two-particle rapidity correlations.

For the equation of state (EOS), we have employed the conformal QGP fluid with the 
thermodynamic pressure $p = \epsilon/3$, and the s95p-PCE EOS \cite{Huovinen:2009yb} 
which is obtained from fits to lattice data for crossover transition and matches 
a realistic hadron resonance gas model at low temperatures $T$,
with partial chemical equilibrium (PCE) of the hadrons at 
temperatures below $T_{\rm PCE} \approx 165$ MeV.
The EOS influences the longitudinal expansion of the fluid and
the two-particle correlations.

In order to obtain the evolution equations for fluctuations, we use
the normalization
$u^\mu u_\mu = (u_0^\mu + \delta u^\mu)(u_{0\mu} + \delta u_{\mu}) = 1$,
orthogonality $\pi^{\mu\nu} u_\nu =0$, and  tracelessness $\pi^\mu_\mu =0$.
It is important to note that noise breaks the boost invariance, as a result
of which the fluctuating quantities depend explicitly on both space-time rapidity
and proper time. Thus the three independent variables are 
$\delta\epsilon \equiv \delta\epsilon(\tau,\eta)$,
$\delta u^\eta \equiv \delta u^\eta(\tau,\eta)$, 
$\delta \pi^{\eta\eta} \equiv \delta \pi^{\eta\eta}(\tau,\eta)
= -(\delta\pi^{xx} + \delta\pi^{yy})/\tau^2$.
Further, since $\Xi^{\mu\nu}$ satisfies the same constraints as 
$\pi^{\mu\nu}$, viz. $u_\mu \Xi^{\mu\nu} =0$ and $\Xi^\mu_\mu =0$, this results
in one independent component, which we take as $\Xi^{\eta\eta}$.
The fluctuating part of the energy-momentum conservation equation (\ref{FCT:eq}) 
can then be written as 
\begin{align} \label{FTevol:eq}
&\frac{\partial}{\partial\tau} (\tau \delta \epsilon) 
+ \frac{\partial}{\partial\eta} (\tau {\cal U}_0 \delta u^\eta) = -\delta {\cal V}, \\ \label{FT1evol:eq}
&\frac{\partial}{\partial\tau}(\tau {\cal U}_0 \delta u^\eta) 
+ \frac{\partial}{\partial\eta} \left(\frac{\cal \delta V}{\tau} \right) 
= - 2 {\cal U}_0 \delta u^\eta , 
\end{align}
where ${\cal U}_0(\tau) \equiv \epsilon_0 + p_0 - \pi_0 = w_0 - \pi_0$ 
depends on the background variables that are functions of proper time only;
$w_0$ is the enthalpy of the fluid. On the other hand,
${\cal \delta V}(\eta,\tau) \equiv \delta p + \tau^2 \delta\pi'^{\eta\eta}$
consists of the fluctuating variables which depend on the space-time 
rapidity as well. We have introduced a stochastic variable
\begin{align}\label{pip:eq}
\delta \pi'^{\eta\eta} = \delta\pi^{\eta\eta} + \Xi^{\eta\eta} 
\equiv  - \delta \pi'/\tau^2,
\end{align}
whose evolution will be derived below.

The stochastic part of the dissipative equation corresponding to MIS or CE, 
can be obtained from Eq. (\ref{IS:eq}) or (\ref{SOSHEAR:eq}) by using 
the linearization Eq. (\ref{nonlin:eq}). For Bjorken expansion, the evolution equation for the 
independent fluctuating component, $\delta\pi'$, reads
\begin{align}\label{Fpievol:eq}
\frac{\partial\delta\pi'}{\partial\tau} + \frac{\delta\pi'}{\tau_\pi} 
=& \frac{1}{\tau_\pi} \left[ \tau^2 \xi^{\eta\eta}
+ \frac{4\eta_v}{3s} \left(s_0\delta\theta + \delta s\theta_0 \right) \right] \nonumber\\
& -\lambda_\pi \left( \theta_0 \delta \pi' + \delta\theta \pi_0 \right)  \nonumber \\
& -\frac{\delta\tau_\pi}{\tau_\pi} 
\left( \lambda_\pi\theta_0\pi_0 + \frac{d\pi_0}{d\tau} \right).
\end{align}
The new noise term $\xi^{\eta\eta}$ defines the equation of motion of $\Xi^{\eta\eta}$, which 
for the MIS equation is
\begin{align}\label{ximis:eq}
\dot \Xi^{\langle\eta\eta\rangle} = - \frac{1}{\tau_\pi} 
\left( \Xi^{\eta\eta} - \xi^{\eta\eta} \right) - \lambda_\pi \Xi^{\eta\eta}\theta ,  
\end{align}
and for the CE equation is
\begin{align}\label{xice:eq}
\dot \Xi^{\langle\eta\eta\rangle} = - \frac{1}{\tau_\pi} 
\left( \Xi^{\eta\eta} - \xi^{\eta\eta} \right)
- \frac{10}{7} \Xi^{\langle\eta}_\gamma \sigma^{\eta \rangle\gamma}
- \lambda_\pi \Xi^{\eta\eta}\theta . 
\end{align}
We recall that in the derivation of Eq. (\ref{Fpievol:eq}), the ratio of the shear viscosity 
and the total entropy density $\eta_v/s$ is kept fixed during the entire evolution.
The local expansion rate of the fluid due to velocity fluctuation is of the form 
$\delta \theta \equiv \partial \cdot \delta u^\eta = \partial_\eta \delta u^\eta$.
The variation of the relaxation time $\tau_\pi$ due to thermal fluctuation is
\begin{align}\label{vartpi:eq}
\delta\tau_\pi = \delta \left( 2\eta_v\beta_2 \right)  
= - \tau_\pi \frac{\delta T}{T_0}.
\end{align}
Equation (\ref{Fpievol:eq}) involves the noise term  $\xi^{\eta\eta}$ that generates the 
fluctuations, which in turn, evolve via the fluctuating hydrodynamic equations.

The equations are closed once the noise $\xi^{\eta\eta}$ (or equivalently $\Xi^{\eta\eta}$) is specified. 
The autocorrelation of $\Xi^{\mu\nu}$ can be obtained using Eq. (\ref{FCT:eq}):
\begin{align}\label{auotocor:eq}
\langle \partial_\mu \Xi^{\mu\nu}(x) \partial_\alpha \Xi^{\alpha\beta}(x') \rangle =&
\langle \partial_\mu \left(-\delta T_{\rm id}^{\mu\nu} - \delta\pi^{\mu\nu}\right)(x) \nonumber\\
&\times 
\partial_\alpha \left(-\delta T_{\rm id}^{\alpha\beta} - \delta\pi^{\alpha\beta}\right)(x') \rangle,
\end{align}
along with the use of modes in the dissipative hydrodynamic equations and
also employing the fluctuation-dissipation theorem \cite{Young:2013fka}. 

Alternatively, the autocorrelations can also be derived from the non-equilibrium
entropy four-current and using the fluctuation-dissipation theorem \cite{Kapusta:2011gt,Kumar:2013twa}. 
In the theory of quasi-stationary fluctuations \cite{Landau},
the rate of change of total entropy can be expressed as $dS/dt = - \sum_a \dot{x_a} X_a$, 
where the generalized forces $X_a = - \partial S/\partial x_a$ are conjugate to the set of
variables $x_a$. For a system close to equilibrium, the evolution of $x_a$ may be
approximated as
\begin{equation}\label{Onsager:eq}
\dot{x}_a = - \sum_b \gamma_{ab}X_b + y_a ,
\end{equation}
where $\gamma_{ab}$ are the Onsager coefficients. 
The random fluctuations $y_a$ then satisfy the autocorrelations 
$\langle y_a(t) y_b(t')\rangle = (\gamma_{ab}+\gamma_{ba})\delta(t-t')$.
In terms of the nonequilibrium part of total energy-momentum tensor,
$\pi'^{\mu\nu} = \pi^{\mu\nu} + \Xi^{\mu\nu}$, 
the entropy four-current (up to second-order) in the MIS and CE theories
can be written as \cite{Chattopadhyay:2014lya}
\begin{equation}\label{S_MIS:eq}
S^{\mu} = s u^{\mu} - \frac{\beta_2}{2T}u^{\mu}\pi'^{\alpha\beta}\pi'_{\alpha\beta},
\end{equation}
where the equilibrium entropy density $s = (\epsilon + p)/T$.
From the total (average plus noise) conservation $\partial_{\mu}T^{\mu\nu} = 0$, one obtains
from Eq. (\ref{S_MIS:eq})
\begin{equation}
\frac{dS}{dt} = \int d^3 \! x \frac{\pi'^{\mu\nu}}{T} \left[ \nabla_{\mu} u_{\nu} - \beta_2
\dot{\pi}'_{\mu\nu} - \beta_2 \lambda_\pi \theta \pi'_{\mu\nu}  \right].
\end{equation}
Identifying 
\begin{align}\label{ide}
\dot x_a \to & \pi'^{\mu\nu} \nonumber \\
X_b \to &  - \frac{1}{T} \left[\nabla_\mu u_\nu - \beta_2\dot{\pi}'_{\mu\nu} 
- \beta_2 \lambda_\pi \theta \pi'_{\mu\nu} \right] \equiv X_{\mu\nu}, \nonumber
\end{align}
and in analogy with Eq. (\ref{Onsager:eq}), one can write
\begin{equation}\label{Landau:eq}
\pi'^{\mu\nu} = - \gamma^{\mu\nu\alpha\beta} X_{\alpha\beta} + \xi^{\mu\nu}.
\end{equation}
Owing to symmetries of $\pi'^{\mu\nu}$, one gets
$\gamma^{\mu\nu\alpha\beta} = \gamma^{\nu\mu\alpha\beta}$, 
$\gamma^{\mu\alpha\beta}_{\mu} = 0$, and 
$\gamma^{\mu\nu\alpha\beta} u_\mu = 0$.
Note that the identification of $X_{\mu\nu}$ is not unique as the
transformation $X_{\mu\nu} \to X_{\mu\nu} + H_{\mu\nu}$, 
keeps $dS/dt$ invariant,
where $H_{\mu\nu}$ is any tensor orthogonal to $\pi'^{\mu\nu}$.
To obtain an autocorrelation which is insensitive to such
transformations, the above constraints for 
$\gamma^{\mu\nu\alpha\beta}$ with respect to $\mu,\nu$ indices
should also follow for the $\alpha, \beta$ indices.

In the MIS theory, we obtain $\gamma^{\mu\nu\alpha\beta}$ by comparing 
Eq. (\ref{IS:eq}) (using $\pi^{\mu\nu} = \pi'^{\mu\nu} - \Xi^{\mu\nu}$) 
and Eq. (\ref{Landau:eq}):
\begin{align}
\gamma^{\mu\nu\alpha\beta} = 2 \eta_v T \Delta^{\mu\nu\alpha\beta}.
\end{align} 
From Eqs. (\ref{Onsager:eq}) and (\ref{Landau:eq}) and using the above
expression of $\gamma^{\mu\nu\alpha\beta}$,
the autocorrelation in the MIS theory can be written as
\begin{align}\label{MInois:eq}
\langle \xi^{\mu\nu}(x) \xi^{\alpha\beta}(x') \rangle 
= 4\eta_v T \Delta^{\mu\nu\alpha\beta} \; \delta^4(x-x').
\end{align}
Similarly, to obtain the autocorrelation in the Chapman-Enskog (CE) theory 
we compare Eq. (\ref{Landau:eq}) with Eq. (\ref{SOSHEAR:eq}) for $\pi'^{\mu\nu}$. 
The $\gamma^{\mu\nu\alpha\beta}$, that is consistent with the 
constraints as stated above, is found to be 
\begin{align}
\gamma^{\mu\nu\alpha\beta} =& 2 \eta_v T \Big( \Delta^{\mu\nu\alpha\beta} 
- \frac{10}{7}\beta_2 \Delta^{\mu\nu}_{\zeta\kappa} \pi'^\zeta_\gamma 
\Delta^{\kappa\gamma\alpha\beta} \nonumber \\ 
& + 2 \tau_\pi \Delta^{\mu\nu}_{\zeta\kappa} \omega^\zeta_\gamma 
\Delta^{\kappa\gamma\alpha\beta} \Big).
\end{align}
Note that the second and third terms in the above equation reproduce 
$\pi'^{\langle\mu}_\gamma \sigma^{\nu\rangle\gamma}$ and
$\omega^{\langle\mu}_\gamma \sigma^{\nu\rangle\gamma}$, respectively,
in the CE theory. Moreover, these terms also give (via contraction with $X_{\alpha\beta}$) 
additional higher order terms which can be neglected in our second-order formalism.
Correspondingly one obtains the noise autocorrelation:
\begin{align}\label{CEnois:eq}
\langle \xi^{\mu\nu}(x) \xi^{\alpha\beta}(x') \rangle 
= & 4\eta_v T \Big( \Delta^{\mu\nu\alpha\beta}  
- \frac{5}{7}\beta_2  \Delta^{\mu\nu}_{\zeta\kappa}\pi'^\zeta_\gamma 
\Delta^{\kappa\gamma\alpha\beta} \nonumber \\ 
& - \frac{5}{7}\beta_2  \Delta^{\alpha\beta}_{\zeta\kappa}\pi'^\zeta_\gamma 
\Delta^{\kappa\gamma\mu\nu} \nonumber \\
& + \omega-{\rm terms} \Big) \delta^4(x-x').
\end{align}
In the boost-invariant case, the autocorrelation for the independent component 
$\xi^{\eta\eta}$ in the MIS (Eq. (\ref{MInois:eq})) and CE (Eq. (\ref{CEnois:eq})) 
dissipative formalisms reduce to 
\begin{align}\label{Bjnoise:eq}
\langle \xi^{\eta\eta}(\eta,\tau) \xi^{\eta\eta}(\eta',\tau') \rangle 
= & \frac{8 \eta_v(\tau) T_0(\tau)}{3 A_\perp \tau^5}
\left[1 - {\cal A}\beta_2\pi_0 \right]  \nonumber \\
& \times \delta(\tau-\tau')\delta(\eta-\eta').
\end{align}
Note that the autocorrelation depends only on the background quantities as we 
have treated the noise as a perturbation on top of background evolution.
The coefficient ${\cal A} = 0$ in the MIS theory, and ${\cal A} = 5/7$ 
in the CE formalism. The delta function in the transverse direction 
$\delta({\bf x} - {\bf x'})\delta({\bf y} - {\bf y'}) = 1/A_\perp$ is represented by the inverse 
of the transverse area $A_\perp$ of the colliding nuclei. The random variable 
$\xi^{\eta\eta}(\eta,\tau)$ is the stochastic source that obeys the energy-momentum conservation,
and propagates each fluctuation $\delta T^{\mu\nu}$ up to later times to their thermal
expectation values. In the Navier-Stokes limit, one can show \cite{Kapusta:2011gt} that
the autocorrelation for $\Xi^{\eta\eta}$ has an identical form 
of Eq. (\ref{Bjnoise:eq}) with ${\cal A} = 0$.
It may be mentioned that the autocorrelation has nonvanishing values
in the transverse directions, 
$\langle \Xi^{ii}(\eta,\tau) \Xi^{ii}(\eta',\tau') \rangle ~(i\equiv x,y)$. 
Consequently, for boost-invariant longitudinal expansion with 
transverse symmetry, a perturbation
generated at any space-time point in the fluid will propagate also in the transverse direction
with the sound velocity of the medium. In the present study, we have ignored such transverse 
motion of these ``ripples" \cite{Shi:2014kta}.
The hydrodynamic fluctuation Eqs. (\ref{FTevol:eq})-(\ref{Fpievol:eq})
are solved perturbatively in the $\tau$-$\eta$ coordinates using the 
MacCormack (a predictor-corrector type) method.

\section{Freeze-out and two-particle rapidity correlations}

The freeze-out of a near-thermalized fluid to a free-streaming (noninteracting) particles 
can be obtained via the standard Cooper-Frye prescription \cite{Cooper:1974mv}. 
For a boost-invariant scenario without fluctuations, freeze-out on a hypersurface 
of constant temperature would be equivalent to freeze-out at a constant proper time.
Inclusion of fluctuation, breaks the boost invariance of the system.
In an event, the total temperature would be the sum of constant background temperature
and the fluctuating temperature which varies for different cells. 
We shall consider freeze-out at a constant background temperature $T_f$ 
so that for any hydrodynamic variable, 
$X(\tau_f,\eta)= X_0(\tau_f) + \delta X (\eta,\tau_f)$, 
the fluctuating field $\delta X(\eta,\tau_f)$ varies 
on the hypersurface; $\tau_f$ is the freeze-out time corresponding to $T_f$. 
For such isothermal (and isochronous) freeze-out at a constant background $T_f$, 
the particle spectrum can be obtained from 
\begin{align}\label{CF:eq}
E\frac{dN}{d^3p} = \frac{g}{(2\pi)^3} \int_\Sigma  d\Sigma_\mu p^\mu f(x,p),
\end{align}
where $p^\mu$ is the four-momentum of the particle with degeneracy $g$ and
$d\Sigma^\mu$ is the outward-directed normal vector on an infinitesimal element 
of the hypersurface $\Sigma(x)$.

In the present ($\tau,x,y,\eta$) coordinate system, the three-dimensional
volume element at freeze-out is  
\begin{align}\label{volel}
d\Sigma_\mu =&  \tau_f \left(\cosh\eta, {\bf 0}, - \sinh\eta \right) d\eta d{\bf x}_\perp.
\end{align}
The four-momentum of the particles is
\begin{align}\label{4mom}
p^\mu = (m_T\cosh y, {\bf p}_T,  m_T\sinh y),
\end{align}
where $m_T = \sqrt{p_T^2 + m^2}$ is the transverse mass of the particle with transverse 
momentum $p_T$ and kinetic rapidity $y =\tanh^{-1}(p_z/p_0)$.  The integration measure 
at the constant temperature freeze-out hypersurface $\Sigma(x)$ is then
$p^\mu d\Sigma_\mu = m_T \: \cosh(y-\eta) \tau_f d\eta d{\bf x}_\perp$.

The phase-space distribution function at freeze-out, 
$f(x,p) = f_{\rm eq}(x,p) + f_{\rm vis}(x,p)$ contains the
 equilibrium contribution 
\begin{align}\label{feq:eq}
f_{\rm eq} = {\rm exp}[p \cdot u/T \pm 1]^{-1} \approx {\rm exp}(-p\cdot u)/T.
\end{align}
The nonequilibrium viscous correction in the MIS theory has the form
based on Grad's 14-moment approximation \cite{Grad}:
\begin{align}\label{fvis_MIS:eq}
f_{\rm vis} = f_{\rm eq}(1 \mp f_{\rm eq}) \frac{p^\mu p^\nu \pi'_{\mu\nu}}{2(\epsilon+p)T^2} 
\approx  f_{\rm eq} \frac{p^\mu p^\nu \pi'_{\mu\nu}}{2(\epsilon+p)T^2}. 
\end{align}
The last expression in Eq. (\ref{fvis_MIS:eq}) is valid in the large temperature limit,
and the total values (noiseless plus noise) for the hydrodynamic variables 
$X \equiv X(\tau_f,\eta)$ are evaluated at the freeze-out hypersurface coordinates.  
In the linearized limit, the total distribution function $f(x,p)$ can be expanded as
\begin{align}\label{f:eq}
f(x,p) = f_0(x,p) + \delta f(x,p).
\end{align}
The noiseless part of the distribution function $f_0(x,p)$ has contributions from ideal 
and viscous fluctuations:
\begin{align}\label{f0:eq}
f_0 =& (f_{\rm eq})_0 
\left( 1 + {K_0}_{\mu \nu} \pi_0^{\mu\nu} \right),
\end{align}
where $K_0^{\mu\nu} = p^\mu p^\nu[2(\epsilon_0+p_0)T_0^2)]^{-1}$, and  
the total temperature $T=T_0 + \delta T$. We recall that $T_0 \equiv T_f$ 
for freeze-out at a constant background temperature.
Similarly, the noise part $\delta f(x,p)$ can be written as the sum of ideal 
and viscous fluctuations. Using the linearization Eq. (\ref{nonlin:eq}), this becomes
\begin{align}\label{f1:eq}
\delta f =& \delta f_{\rm eq} + {K_0}_{\mu\nu} \Big[ 
\delta f_{\rm eq} \: \pi_0^{\mu\nu} + (f_{\rm eq})_0 \: \delta \pi'^{\mu\nu} \nonumber \\
& - (f_{\rm eq})_0 \: \pi_0^{\mu\nu}
\left( 2\frac{\delta T}{T_0} + \frac{\delta \epsilon+\delta p}{\epsilon_0+p_0} \right)  \Big] ,
\end{align}
where $(f_{\rm eq})_0 = {\rm exp}(-u_0^\mu p_\mu/T_0)$ 
and $\delta f_{\rm eq} = (f_{\rm eq})_0 (\delta T u_0^\mu p_\mu/T_0^2 - \delta u^\mu p_\mu/T_0)$
are, respectively, the noiseless and the noise parts of the ideal 
distribution function, and the terms inside the square brackets 
in Eq. (\ref{f1:eq}) stem from viscous fluctuations.

The rapidity distribution of the particle, corresponding to Eq. (\ref{CF:eq}), then reduces to 
\begin{align}\label{rapidity:eq}
\frac{dN}{dy} =& \frac{g\tau_f A_\perp}{(2\pi)^3} \int d\eta \: \cosh(y-\eta)  \nonumber\\
& \times \int dp_x dp_y \: m_T [f_0(x,p) + \delta f(x,p)] \nonumber\\
& \equiv (dN/dy)_0  + \delta (dN/dy) .
\end{align}
Here $A_\perp = \int d{\bf x}_\perp$ is the usual transverse area of Eq. (\ref{Bjnoise:eq}). 
For the boost-invariant longitudinal flow, the particle rapidity distribution 
of the average part can be written as 
\begin{align}\label{rap0:eq}
\left(\frac{dN}{dy}\right)_{\!\! 0}  =& ~\frac{g\tau_f T_0^3 A_\perp}{(2\pi)^2} 
\int \frac{d\eta}{\cosh^2(y-\eta)} \: 
\Big[ \Gamma_3(y-\eta) \nonumber \nonumber \\
& + \frac{\pi_0}{4 w_0} 
\left({\cal C}(y-\eta) \Gamma_5(y-\eta) 
 - \frac{m^2}{T_0^2}\Gamma_3(y-\eta) \right) \Big].
\end{align}
Here $\Gamma_k(x) \equiv \Gamma(k, m\cosh x /T_0)$ denotes the incomplete Gamma function 
of the $k$th kind \cite{Stegun} and ${\cal C}(x) = 3~{\rm sech}^2x -2$.
The second term within the brackets corresponds to viscous corrections. 
The fluctuating parts can be expressed as 
\begin{align}\label{rapf:eq}
\delta\frac{dN}{dy} &= ~\frac{g\tau_f T_0^3 A_\perp}{(2\pi)^2} 
\int  d\eta \: \Big[ {\cal F}_T (y-\eta) \frac{\delta T(\eta)}{T_0}  \nonumber\\
&+ {\cal F}_u (y-\eta) \tau_f \delta u^\eta(\eta) + {\cal F}_\pi (y-\eta) 
\frac{\delta\pi'(\eta)}{w_0} \Big].
\end{align}
Here ${\cal F}_{T,u,\pi}$ are the coefficients of the fluctuations, 
($\delta T, \delta u^\eta, \delta \pi'$), that are obtained by performing the momentum integrals.
In the MIS theory these are given by
\begin{align}\label{Coef_Tup:eq}
{\cal F}_T \cosh^2 x  =&  \Gamma_4(x) 
- \frac{\pi_0}{4w_0} \Big[ \frac{m^2}{T_0^2}\left(\Gamma_4(x) - \kappa\Gamma_3(x) \right) \nonumber\\ 
& - {\cal C}(x) \left(\Gamma_6(x) - \kappa\Gamma_5(x) \right) \Big], \\
{\cal F}_u \cosh^2 x  =&   \tanh x ~\Gamma_4(x)  
- \frac{\pi_0}{4w_0} \tanh x  ~\Big[ \frac{m^2}{T_0^2}\Gamma_4(x) \nonumber \\
& - {\cal C}(x)\Gamma_6(x) - 4\Gamma_5(x) \Big], \\ 
{\cal F}_\pi \cosh^2 x  =& \frac{1}{4} 
\Big[{\cal C}(x) \Gamma_5(x) - \frac{m^2}{T_0^2}\Gamma_3(x) \Big],
\end{align}
where $\kappa = 2 + (T_0/w_0)\partial w_0 /\partial T_0$.
The two-particle rapidity correlator due to fluctuations can then
be written as 
\begin{align}\label{FRapCor:eq}
\left\langle \delta\frac{dN}{dy_1} \ \delta\frac{dN}{dy_2} \right\rangle =&  
\left[ \frac{g\tau_f T_0^3 A_\perp}{(2\pi)^2} \right]^2  
\int \! d\eta_1 \int \! d\eta_2  \nonumber \\
& \times \sum_{X,Y} {\cal F}_X (y_1-\eta_1){\cal F}_Y (y_2-\eta_2) \nonumber\\
& \times \langle X(\eta_1) Y(\eta_2) \rangle .
\end{align}
Here $(X,Y) \equiv (\delta T, \delta u^\eta, \delta \pi')$ and $\langle X(\eta_1) Y(\eta_2) \rangle$ 
are the two-point correlators between the fluctuating variables calculated at the freeze-out hypersurface.

In the Chapman-Enskog-like approach of iteratively solving Boltzmann equation, the
viscous correction in the nonequilibrium distribution function has the form 
\cite{Jaiswal:2013npa,Bhalerao:2013pza} 
\begin{align}\label{fvis_CE:eq}
f_{\rm vis} \approx  f_{\rm eq} \frac{5 p^\mu p^\nu \pi'_{\mu\nu}}{8pT(u.p)}, 
\end{align}
with a total flow velocity $u^\mu \equiv u^\mu(\tau_f,\eta)$. Following 
similar procedure as done in the MIS theory, the two-particle rapidity correlations
in the CE formalism give the same form as in Eq. (\ref{FRapCor:eq}) but with modified
coefficients ${\cal F}_{T,u,\pi}$.

\section{Results and discussions}

We shall explore two-particle rapidity correlations induced by thermal fluctuations 
in the Bjorken expansion. A clear understanding of this can be achieved 
by calculating the time evolution of the correlations
among the hydrodynamic variables. The fluctuations employed in our study generates 
singularities in the correlators at zero separation in rapidity and at the sound 
horizons that corresponds to maximum distance propagated by the sound wave along rapidity.
In the Navier-Stokes theory, the singular and regular parts of the correlators can 
be obtained analytically \cite{Kapusta:2011gt}; see Appendix A.
Figure 1 shows the rapidity dependence of these equal-time correlators for inviscid fluid 
$\langle X(\tau,\eta_1) Y(\tau,\eta_2) \rangle$ with $(X,Y) \equiv (\delta T, \delta u^\eta)$.
Shear viscosity is neglected in the evolution but
accounted for in the noise correlator $\Xi^{\eta\eta}$ of Eq. (\ref{Bjnoise:eq}).
These correlators represent a wake of the medium behind the shock front associated 
with the noise propagation.
The regular part of the temperature-temperature correlator, $\langle \delta T \delta T \rangle$,
peaks at zero separation due to short-range correlation which builds up rapidly with 
increasing proper time. At later times, the peak value decreases and the correlator spreads 
farther in rapidity due to expansion of the fluid. 
The time evolution of these structures in the Bjorken expansion is to be contrasted 
with that of a uniform static system; see Appendix B.
The equal-time long-range correlations in the static fluid vanish and correlations 
at later times are due to propagation of sound waves in  
opposite directions. This clearly underscores the importance of underlying background
flow that influences the propagation of fluctuations.

\begin{figure}[t]
\includegraphics[width=\linewidth,height=4.0in]{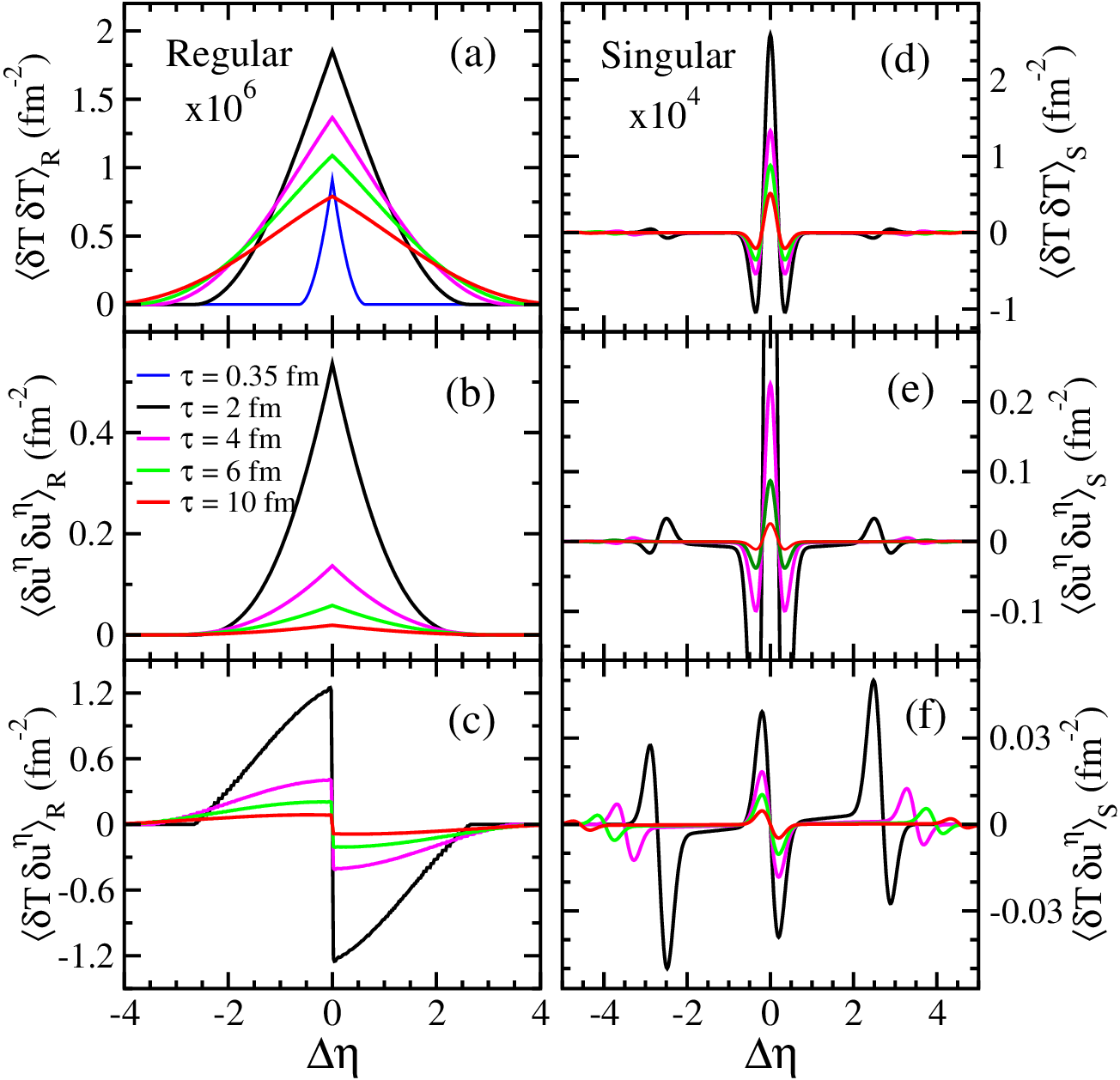}
\caption{Regular and singular parts of the equal time 
temperature-temperature, velocity-velocity and temperature-velocity correlators computed 
as a function of space-time rapidity difference $\Delta\eta$ at various proper times in
the Navier-Stokes theory. The results are for ultrarelativistic gas EOS 
($p=\epsilon/3$) with initial temperature of $T_0 = 550$ MeV, 
proper time $\tau_0 = 0.2$ fm/c. The shear viscosity to entropy density ratio of 
$\eta_v/s = 1/4\pi$ is accounted only in the noise correlator.}
\label{fig:Corr}
\end{figure}

In Fig. \ref{fig:Corr}, we also show the regular part of the velocity-velocity correlator, 
$\langle \delta u^\eta \delta u^\eta \rangle$. It exhibits a similar 
rapidity dependence as seen in $\langle \delta T \delta T \rangle$,
however with a much smaller magnitude. In contrast, the regular part of the
temperature-velocity correlator, $\langle \delta T \delta u^\eta \rangle$
is an odd function in $\Delta\eta$. As a consequence this correlator vanishes at 
$\Delta\eta =0$ and turns negative (positive) for positive (negative) values 
of rapidity separation. Note that the ``cross" correlators follow 
$\langle \delta T \delta u^\eta \rangle =-\langle \delta u^\eta \delta T\rangle$. 

Analytic results for the singular part of the correlator can be obtained in the
Navier-Stokes theory for an ultra-relativistic gas EOS, see Appendix A.
Figure 1 shows the singular part of the equal-time correlators 
wherein the theta function and its higher-order derivatives have been smeared using a 
normalized Gaussian distribution of width $\sigma_\eta =0.2$. In all these correlators, 
the singularities at $\Delta\eta = 0$ arise from self-correlations, and those at large rapidity 
separations are induced by sound horizons at $\Delta\eta = \pm 2c_s\log(\tau/\tau_0)$.
We note that inclusion of viscosity would dampen the singularities and thereby smear the 
structures in the longitudinal correlations. 

As analytic solutions for thermal fluctuations in second-order dissipative hydrodynamics do not
exist, the singular and regular parts of the correlators cannot be separated. However, in the
numerical simulation of noise in second-order hydrodynamics, the smearing functions ${\cal F}_X$ 
in Eq. (\ref{FRapCor:eq}) smoothen out all the singularities 
in the total correlation function $\langle X(\tau,\eta_1) Y(\tau,\eta_2) \rangle$.
Hence the computed two-particle rapidity correlation at freeze-out would show a clear structure. 

\begin{figure}[t]
\includegraphics[width=\linewidth]{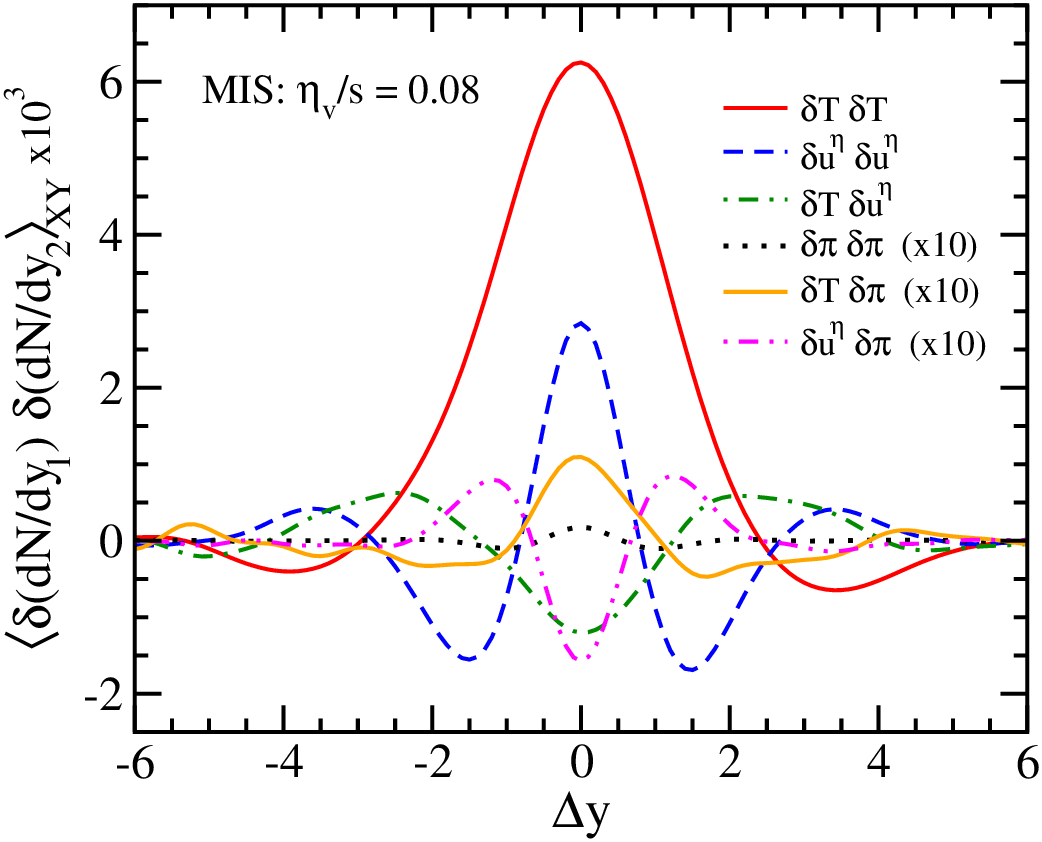}
\caption{Two-particle rapidity correlations from various fluctuations calculated at 
freeze-out for charged pions in the M\"uller-Israel-Stewart (MIS) dissipative hydrodynamics. 
The results are for ideal gas EOS ($p=\epsilon/3$) with initial temperature $T_0 = 550$ MeV, 
proper time $\tau_0 = 0.2$ fm/c, freeze-out temperature $T_f = 150$ MeV and shear viscosity 
to entropy density ratio $\eta_v/s = 1/4\pi$.} 
\label{fig:FFCor}
\end{figure}

In the second-order M\"uller-Israel-Stewart viscous hydrodynamics, 
we present in Fig. \ref{fig:FFCor} the various components, $X,Y \equiv \delta T, \delta u^\eta, \delta \pi$,
of the rapidity correlators $\langle (\delta dN/dy_1) (\delta dN/dy_2) \rangle_{X,Y}$
(of Eq. (\ref{FRapCor:eq})) for charged pions as a function of kinematic rapidity separation $\Delta y = y_1-y_2$.
This has been obtained by convoluting the two-point correlators $\langle X(\tau,y_1) Y(\tau,y_2) \rangle$,
at freeze-out with the respective smearing functions ${\cal F}_{X,Y}$. 
The calculations are for initial values of temperature $T_0 = 550$ MeV, proper 
time $\tau_0 = 0.2$ fm/c and the freeze-out temperature is taken as $T_f = 150$ MeV.
A constant $\eta_v/s = 0.08$ is used in both the average and noise parts of the 
evolution equations. The two-particle correlation functions
are essentially manifestations of the sum of their regular and singular parts;
see Fig. \ref{fig:Corr} for the correlators in the Navier-Stokes case. 
The smearing functions, namely, $F_{\delta T}$ (which is Gaussian about $\delta\eta=0$) and
$F_{\delta u^\eta}$ (which peaks at $\Delta\eta \simeq \pm 1.5$ and vanishes at $\Delta\eta =0$) 
broadens these correlators when convoluted. While the peak at $\delta\eta=0$ is dominated by the 
temperature-temperature correlation function, the structures seen at 
$\Delta\eta \simeq 2-4$ for the $\delta T \delta T$,  $\delta u^\eta  \delta u^\eta$ 
and $\delta T \delta u^\eta$ correlations are similar in magnitude but have distinct 
rapidity dependence. The contributions to the correlation functions involving the 
viscous stress tensor $\delta\pi$ are found to be much smaller.

\begin{figure}[t]
\includegraphics[width=\linewidth]{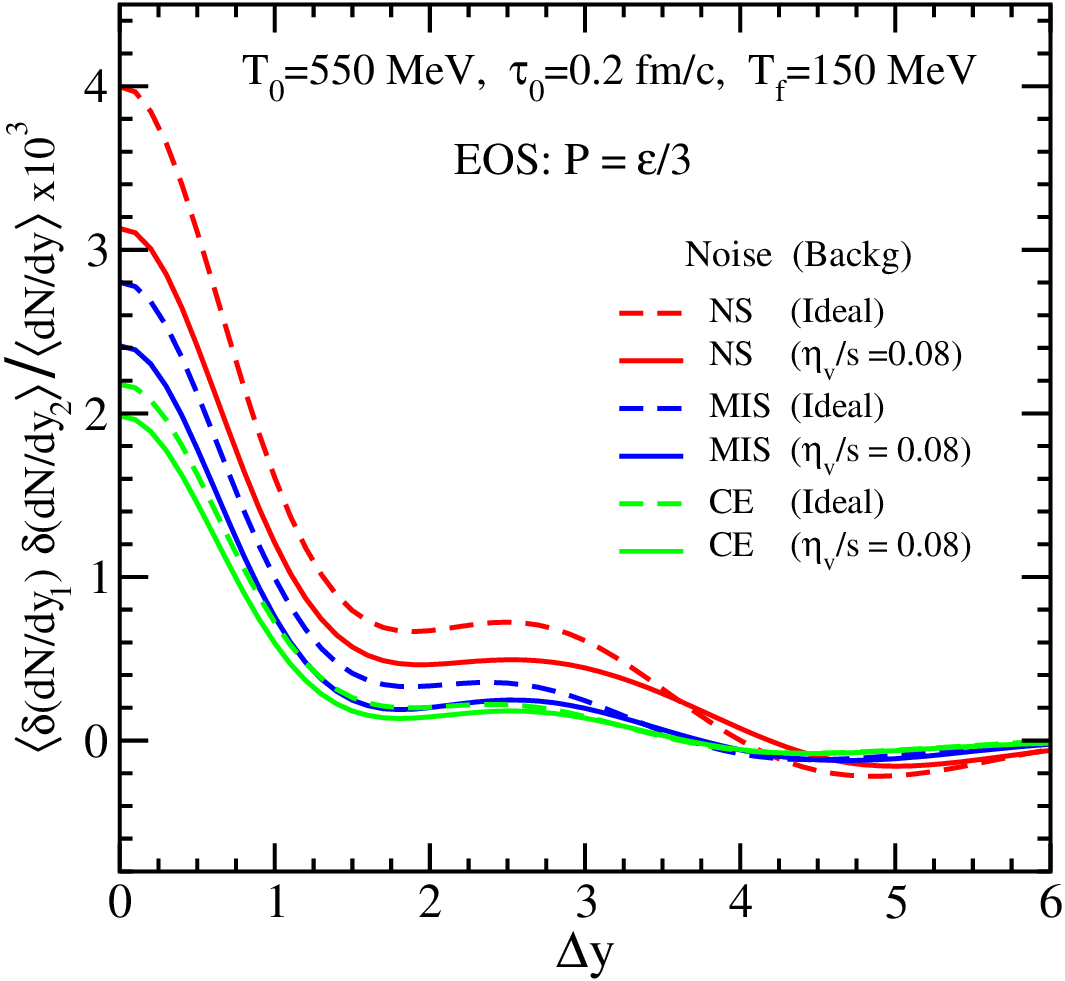}
\caption{Correlation function for charged pions normalized with the single-particle 
rapidity distribution in the Navier-Stokes (NS), M\"uller-Israel-Stewart (MIS), 
and Chapman-Enskog (CE) formalisms for the evolution of thermal noise and compared with
the ideal background (noiseless) hydrodynamic evolution. 
The initial and freeze-out conditions are same as in Fig. \ref{fig:FFCor}.}
\label{fig:N12}
\end{figure}

In Fig. \ref{fig:N12} we compare the total two-particle rapidity correlation for charged pions for the
Navier-Stokes, M\"uller-Israel-Stewart and Chapman-Enskog viscous evolutions for an ultra-relativistic gas EOS 
($p=\epsilon/3$). The initial and freeze-out conditions are the same as in Fig. \ref{fig:FFCor}.
We shall first consider the case of ideal hydrodynamics for the background evolution 
and explore various dissipative equations for the evolution of thermal fluctuations. 
The fluctuation in the Navier-Stokes theory gives rise to a larger peak at small
rapidity separations as compared to that in the second-order viscous evolutions. 
This is mainly due to faster build-up of all the correlators and in particular the 
dominant temperature-temperature correlations in the first-order viscous evolution.
In the Chapman-Enskog case, the correlation strength at $\Delta y \approx 0$ is smallest 
due to the larger coefficient $\lambda_\pi =38/21$, that
results in a slower approach of the viscous fluctuations towards the Navier-Stokes limit.

The inclusion of viscosity in the background evolution damps the correlation peak for
all the cases studied. As expected, the maximum reduction in correlation strength occurs
in the first-order theory. It may be mentioned that previous studies of rapidity correlations
ignored the variation of the relaxation time $\delta \tau_\pi$ of Eq. (\ref{vartpi:eq})
\cite{Kapusta:2011gt,Young:2013fka,Young:2014pka}. We have found that such an assumption
is justified as the rapidity correlation remains practically unaltered when thermal
fluctuation of $\tau_\pi$ was not considered.

\begin{figure}[t]
\includegraphics[width=\linewidth]{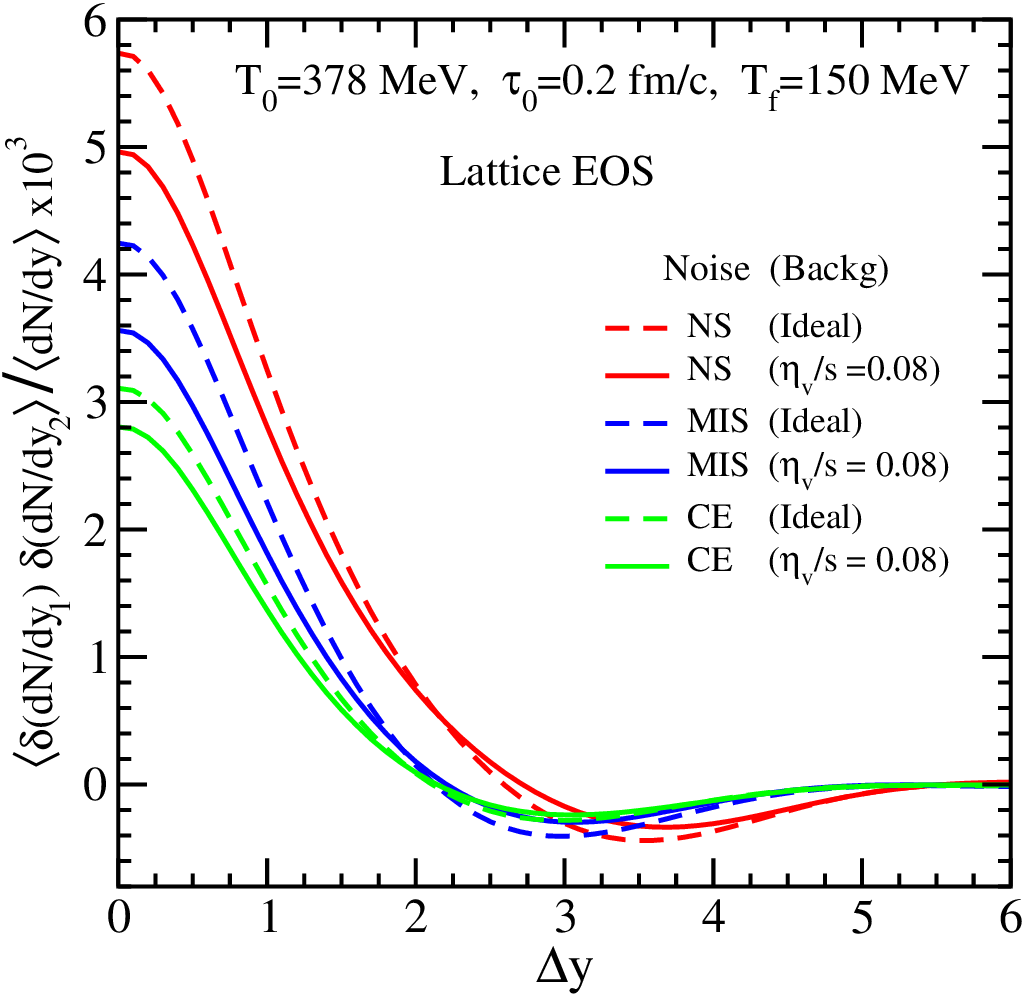}
\caption{Similar to Fig. \ref{fig:N12} but with lattice QCD EOS; see text for details.}
\label{fig:N12_lat}
\end{figure}

Figure \ref{fig:N12_lat} shows numerical results for two-particle rapidity correlation of
charged pions in the NS, MIS, and CE formalisms for the lattice QCD EOS that incorporates 
the transition to a hadron resonance gas at $T_{\rm PCE} \approx 165$ MeV
\cite{Huovinen:2009yb}. It may be mentioned that analytical results for the correlations
cannot be obtained for the lattice EOS even in the Navier-Stokes limit.
While the initial time and freeze-out temperature 
are considered the same as used for ideal gas
EOS, the initial temperature is set at $T_0 = 378$ MeV. This choice stems from the
consideration that the event-averaged single particle rapidity distribution for direct charged pions, 
$\langle dN/dy \rangle$, in this case is practically identical to that in the ideal gas EOS.
Moreover, the freeze-out times
for the lattice and conformal equation of states are found similar for each of the dissipative theories. 
We find that the magnitudes of the correlation between the particles
are enhanced for all the cases in the lattice EOS as compared to that for ideal gas EOS.
This can be understood as due to smaller sound velocity of the fluid near 
the critical temperature $T_c$ which slows the fluid expansion. Consequently, the
correlation is solely from the short-range temperature-temperature correlator and the 
structures associated with the velocity and shear pressure correlators are largely damped
and do not spread in rapidity.

\section{Summary and Conclusions}

We have studied the evolution of thermal fluctuations within relativistic second-order 
dissipative hydrodynamics. The fluctuations were treated in the linearized hydrodynamic 
framework as a perturbation on top of boost-invariant longitudinal expansion of matter. 
The analytic form of the autocorrelation function
was found to be identical for the acausal Navier-Stokes and the causal M\"uller-Israel-Stewart 
theories. However, for the Chapman-Enskog-like dissipative equations, the correlation has an explicit
dependence on the shear stress tensor.  Within the analytically solvable Navier-Stokes limit 
in the Bjorken scenario, we demonstrated that the two-particle rapidity correlation at small rapidity 
separation, $\Delta y \lesssim 2$, is mostly due to temperature-temperature correlations and 
structures seen in the correlations at $\Delta y \approx 2-4$, are caused by varying contributions
involving fluid velocity and shear pressure tensor correlations. 
In general, the two-particle rapidity correlations produced from thermal fluctuations
were found to spread to large distance in rapidity separation with magnitude (and pattern)
that can be well measured in relativistic heavy-ion collisions.
While viscous damping of the correlation is at most $\sim 20\%$,
there is further significant damping at small $\Delta y$, if one goes
from the first-order Navier-Stokes theory to a second-order
dissipative hydrodynamic formulation. As compared to the conformal equation of state,
the softer lattice QCD EOS, with smaller sound velocity, causes reduced propagation of the
fluctuations but leads to a pronounced peak in the rapidity correlations.

\begin{acknowledgments}
We thank Ananta Mishra for useful discussions during the initial stages of the work. 
We also thank Jean-Yves Ollitrault for illuminating discussions.
\end{acknowledgments}


\appendix

\section{Singular part of the correlators in Navier-Stokes theory}

The correlation functions for the fluctuating quantities $(X,Y) \equiv (\delta\epsilon, \delta u^\eta)$, which
are linear functionals of the noise $\Xi^{\eta\eta}$, can be written as 
\begin{align}\label{G:eq}
\langle X(\eta,\tau)  Y(\eta',\tau) \rangle = \frac{2}{A_\perp} \int^{\tau}_{\tau_0} 
\frac{d\tau'}{\tau'^3} \frac{4\eta_v}{3s w_0(\tau')} G_{XY}(\eta-\eta';\tau,\tau') .
\end{align}
These Green functions $G_{XY}(\eta-\eta';\tau,\tau')$ have singular and regular parts
which are obtained from the fluctuation evolution equations.
The fluctuating component $\delta \pi$ of Eq. (\ref{IS:eq}), 
in the Navier-Stokes limit reduces to 
\begin{align}\label{NSpiev:eq}
\delta \pi = \frac{4\eta_v}{3s} \left( s_0 \delta\theta + \delta s \theta_0 \right).
\end{align}
Using this $\delta\pi$ along with the noiseless $\pi_0 = 4\eta_v\theta_0/3$ for the
Navier-Stokes case, the fluctuating quantities 
$\delta\epsilon, \delta u^\eta$ are found from the linearized 
evolution Eqs. (\ref{FTevol:eq}), (\ref{FT1evol:eq}) with coefficients 
${\cal U}_0 = w_0 - (4\eta_v/3s) s_0/\tau$ and 
$\delta{\cal V} = \delta p - \tau^2\Xi^{\eta\eta} 
- (4\eta_v/3s) (s_0 \delta\theta + \delta s\theta_0)$.
In the conformal case, these linearized equations have been solved by Fourier transform of
$\delta\epsilon, \delta u^\eta$ and finding the corresponding Green functions 
$G_{\delta\epsilon}(k;\tau,\tau')$ and $G_{\delta u^\eta}(k;\tau,\tau')$ \cite{Kapusta:2011gt} 
\begin{align}
G_{XY} (\eta-\eta';\tau,\tau') = \int_{-\infty}^{\infty} \frac{d\,k}{2\pi} \, 
e^{ik(\eta-\eta')} \, \tilde{G}_{XY}(k; \tau, \tau'),
\end{align}
where $\tilde{G}_{XY}(k;\tau,\tau') \equiv \tilde{G}_X(k;\tau,\tau') \, \tilde{G}_Y(-k;\tau,\tau')$.
The terms which give rise to singular behavior of $G_{XY}(\eta-\eta';\tau,\tau')$ can be obtained 
analytically by Laurent series expansion of $\tilde{G}_{XY}(k;\tau,\tau')$ in powers of $1/k$. 

\begin{widetext}
Denoting $\rho \equiv 3\delta e/(4e_0)$ and $\omega \equiv \tau\delta u^\eta$, 
the expression for the singular part of $G_{\rho\rho}(\eta-\eta';\tau,\tau')$ stems from
\begin{align}
\tilde{G}_{\rho\rho}^{\mathrm{sing}}(k;\tau,\tau') = (a_1 k^2 + b_1) + (a_2 k^2 + b_2) \cos(2 c_s \gamma k) 
+\frac{a_3 k^2 + b_3}{k} \sin(2 c_s \gamma k).
\end{align}
The coefficients are found to be
\begin{align}
a_1 =& \frac{\beta}{2 c_s^2}, ~~
a_2 = - a1, ~~
a_3 = \beta \left(\frac{1}{c_s} - \frac{\gamma\delta}{2c_s} \right), ~~
b_1 = \beta \frac{c_s^2 + 2\alpha + \delta}{2 c_s^2}, ~~
b_2 = \beta \left( \frac{1}{2} - \frac{2\alpha+\delta}{2 c_s^2} + \frac{\gamma^2\delta^2}{4} 
       - \gamma \delta \right),  \nonumber \\
b_3 =& \beta\left[ \frac{\gamma\delta}{2c_s} 
 \left(c_s^2 - 2\alpha - \delta \right) -  \frac{\gamma\delta^2}{8c_s} 
 - \frac{c_s \gamma^2 \delta^2}{2} + \frac{c_s \gamma^3 \delta^3}{12} + \frac{2\alpha + \delta}{2c_s} \right], 
 \end{align}
where $\alpha = (1-c_s^2)/2$, $\beta = (\tau'/\tau)^{2\alpha}$, 
$\gamma \equiv \log(\tau/\tau')$, and $\delta = \alpha^2/c_s^2$.
The singular behavior of $G_{\rho\omega}(\eta-\eta';\tau,\tau')$:
 \begin{align}
 \tilde{G}_{\rho\omega}^{\mathrm{sing}}(k;\tau,\tau') = d_1 k + d_2 k \, \cos(2 c_s \gamma k) 
  +( d_3 k^2 + d_4) \sin(2 c_s \gamma k),
 \end{align}
where the corresponding coefficients are
\begin{align}
 d_1 =& -\frac{i}{2} \beta \left( 1 - \frac{\alpha + c_s^2}{c_s^2} \right), ~~
 d_2 = \frac{i}{2} \beta \left(\gamma\delta - 1 - \frac{\alpha + c_s^2}{c_s^2} \right), ~~
 d_3 = - \frac{i}{2} \beta \frac{1}{c_s}, \nonumber \\
 d_4  =& - \frac{i}{2} \beta
\left[ -\frac{\alpha + c_s^2}{c_s}\left( 1 - \frac{\gamma \alpha^2}{c_s^2} \right)
    + \frac{\alpha}{c_s}\left( 1 + \frac{\alpha}{2 c_s^2} \right) 
+ \frac{\gamma \alpha^2}{c_s} \left( 1 - \frac{\gamma\alpha^2}{2c_s^2} \right) \right].  
\end{align}
Finally, the singular behavior of $G_{\omega\omega}(\eta-\eta';\tau,\tau')$ originates from:
\begin{align}
\tilde{G}_{\omega\omega}^{\mathrm{sing}}(k;\tau,\tau') = (w_1 k^2 + w_2) + (w_3 k^2 + w_4) \cos(2 c_s \gamma k)
+ \left(\frac{ w_5 k^2 + w_6  }{k} \right) \sin(2 c_s \gamma k),
\end{align}
with coefficients 
\begin{align}
 w_1 =& \frac{1}{2} \beta ,~~
 w_2 = \frac{1}{2} \beta \left( \frac{\alpha + c_s^2}{c_s} \right)^2, ~~
 w_3 = \frac{1}{2} \beta, ~~
 w_4  = - \beta \left[ \frac{\gamma \alpha^2}{c_s} 
        \left( \frac{\gamma\alpha^2}{4c_s} -  \frac{\alpha + c_s^2}{c_s} \right)
        + \frac{1}{2}\left(\frac{\alpha + c_s^2}{c_s} \right)^2 \right],\nonumber \\
 w_5 =& \beta \left[ \frac{\gamma\alpha^2}{2c_s} - \frac{\alpha + c_s^2}{c_s} \right], ~~ 
 w_6 = \beta \left[ \frac{\gamma\alpha^2}{2c_s} 
        \frac{\alpha+c_s^2}{c_s} \left( \frac{\gamma\alpha^2}{c_s} - \frac{\alpha+c_s^2}{c_s} \right)
        + \frac{\gamma \alpha^4}{8 c_s^3} \left( 1 - \frac{2 \gamma^2 \alpha^2}{3} \right) 
        - \frac{\alpha + c_s^2}{c_s} \frac{\delta}{2} \right] . 
\end{align}
\end{widetext}


\section{Thermal fluctuations in a static fluid}

\begin{figure}[t]
\includegraphics[width=\linewidth]{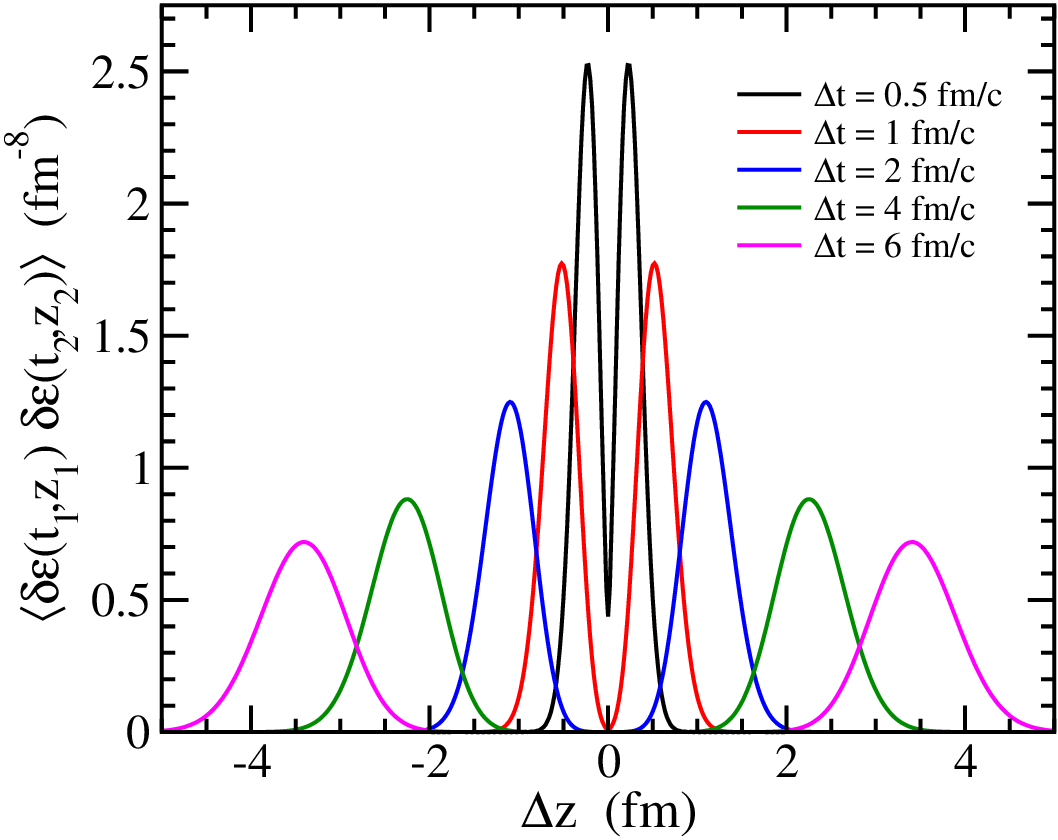}
\caption{Spatial dependence of energy-energy correlations at various time intervals $\Delta t$
due to thermal fluctuations created in a static medium.
The results are for an ultra-relativistic gas EOS with a temperature of $T_0 = 550$ MeV
and shear viscosity to entropy density ratio $\eta_v/s = 1/4\pi$.}
\label{fig:Corr_stat}
\end{figure}

Consider a static uniform fluid in Cartesian coordinates ($t, z$) with fluctuations that depend
on $z$ and are independent of the transverse ($x,y$) directions. 
In the Navier-Stokes theory for a conformal equation of state, the evolution equation for longitudinal 
fluctuations ($\delta\epsilon, \delta u^z$) can be written as 
\begin{align}\label{statevol:eq}
&\frac{\partial\delta \epsilon}{\partial t} 
+ w_0 \frac{\partial \delta u^z}{\partial z} = 0,  \nonumber \\
&w_0 \frac{\partial\delta u^z}{\partial t}
+ c_s^2 \frac{\partial \delta\epsilon}{\partial z} 
- \frac{4}{3}\eta_v \frac{\partial^2 \delta u^z}{\partial z^2} 
+ \frac{\partial\Xi^{zz}}{\partial z} = 0,
\end{align}
where the constant enthalpy of the background is denoted by $w_0 = \epsilon_0 + p_0$. 
The noise correlator takes the form
\begin{align}\label{statnoise:eq}
\langle \Xi^{zz}(t,z) \Xi^{zz}(t',z') \rangle 
=\frac{8\eta_v T_0}{3 A_\perp}\delta(t-t')\delta(z-z').
\end{align}
Equations (\ref{statevol:eq}) can be solved by taking the Fourier transform
\begin{align}\label{statFT:eq}
\delta X(t,z) = \int \frac{d\omega \: dk}{(2\pi)^2} \: e^{-i\omega t} e^{-ikz}  
\delta \tilde{X} (\omega,k),
\end{align}
where the fluctuations are denoted by $X \equiv (\delta\epsilon, \delta u^z)$.
The two-point energy correlator becomes
\begin{align}\label{statcor:eq}
\langle \delta\epsilon(t,z) \delta\epsilon(t',z') \rangle = & \frac{8T_0 \eta_v}{3 A_\perp} 
\int \frac{d\omega \: dk}{(2\pi)^2} \:
e^{-i\omega(t-t')} e^{-ik(z-z')}  \nonumber \\
& \times \frac{k^4}{(\omega^2-c_s^2 k^2)^2 + \alpha^2 k^4 \omega^2} ,
\end{align}
where $\alpha = 4\eta_v/(3w_0)$. For equal-times, the energy correlation becomes 
$\langle \delta\epsilon(t,z) \delta\epsilon(t,z') \rangle = w_0T_0\delta(z-z')/(c_s^2 A_\perp)$.
Thus in a static fluid noise produces only local correlations
and does not induce any long-range structures. For unequal times, the
correlators in Eq. (\ref{statcor:eq}) admit analytic
solutions only in the limit of $\eta_v \to 0$:
\begin{align}\label{statneq:eq}
\langle \delta\epsilon(t,z) \delta\epsilon(t',z') \rangle = \frac{w_0 T_0}{2 c_s^2 A_\perp} 
\Big[& \delta(\Delta z -c_s \Delta t) \nonumber \\
& + \delta(\Delta z + c_s \Delta t) \Big],
\end{align}
where $\Delta t = (t-t')$ and $\Delta z = (z-z')$. Thus one finds that in a static fluid, when 
shear viscosity is neglected in the evolution of fluctuations, the correlations
are produced solely by sound waves of velocity $c_s^2 = \partial p/\partial e$ which
propagate without attenuation. In presence of viscosity, the energy-energy correlator 
of Eq. (\ref{statcor:eq}) has a singular part given by 
$w_0 T_0/(c_s^2 A_\perp){\rm exp}(-c_s^2 \Delta t/\alpha)\delta(\Delta z)$. 
On the other hand, the regular part has to be computed numerically.
The time dependence of regular part of this correlation is shown in Fig. \ref{fig:Corr_stat} 
for a constant background temperature of $T_0 = 550$ MeV and $\eta_v/s = 1/4\pi$. 
With increasing time $\Delta t$, viscosity is seen to reduce the amplitude of the two peaks 
formed at $\Delta z = \pm c_s \Delta t$ as well as broaden the correlations.


\end{document}